# RARE KAON DECAYS AT KTEV

**Marj Corcoran (for the KTeV Collaboration)**
Invited talk at the Workshop on
$e^+e^-$ Physics in the 1-2 GeV Range
Alghero, Italy, September 2003.

*Abstract*

The KTeV experiment at Fermilab has studied a wide range of rare $K_L$ decays. In this talk I focus on the electromagnetic decays $K_L \to \gamma^*\gamma^{(*)}$ where one or both photons is virtual. Recent results are also presented for searches for the CP-violating decay $K_L \to \pi^0 e^+ e^-$ and lepton-flavor violating modes.

## OVERVIEW

The KTeV experiment at Fermilab has studied a wide range of CP-violating and rare $K_L$ decays. KTeV is one of two experiments to measure the direct-CP violating parameter $Re(\epsilon'/\epsilon)$; KTeV has the best measurement of the $K_{e3}$ charge asymmetry $\delta$; KTeV was the first experiment to observe the CP-violating decay $K_L \to \pi^+\pi^-e^+e^-$; KTeV has the best limits on decays with large direct-CP violating contributions: $K_L \to \pi^0\nu\bar{\nu}$, $K_L \to \pi^0 e^+e^-$, and $K_L \to \pi^0\mu^+\mu^-$. In this talk I will concentrate on the electromagnetic modes $K_L \to \gamma^*\gamma^{(*)}$, where either one or both final-state photons are virtual. I will also show a recent limit for $K_L \to \pi^0 e^+e^-$, and discuss lepton-flavor-violation searches.

## FORM FACTORS FOR $K_L \to \gamma^*\gamma^{(*)}$

The electromagnetic modes are interesting in part because of their connection to the decay $K_L \to \mu\mu$. This latter decay can shed light on the CKM matrix element $\rho$ [1], but only if the long distance contributions to the decay are well-understood [2]. Figure 1 shows examples of both the long-distance and short-distance contributions to $K_L \to \mu^+\mu^-$.

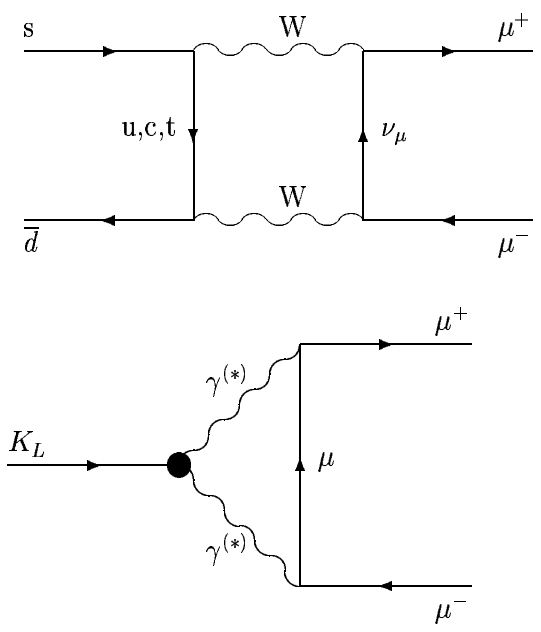

Figure 1. Short distance (top) and long distance (bottom) contributions to $K_L \to \mu^+\mu^-$.

The key to unraveling the long distance contribution is the $K \to \gamma^*\gamma^{(*)}$ form factor. There are two different parameterizations commonly in use, and in general we fit to both parameterizations in order to facilitate comparison to other experiments. One form, due to Bergström, Massó, and Singer (BMS) [3], has two contributions, one from a pseudoscalar-vector-photon vertex ($A_1$), and the second from a pseudoscalar-vector-vector vertex ($A_K$). This form factor, which is applicable only to the $\gamma\gamma^*$ final state, has one free parameter, $\alpha_{K^*}$:

$$f(q^2) = A_1(q^2) + \alpha_{K^*} A_K(q^2)$$

where $q^2$ refers to the 4-momentum squared of the one virtual photon.

The second widely-used $K \to \gamma^*\gamma^*$ form factor, due to D'Ambrosio, Isidori, and Portolés (DIP), is compatible with a chiral perturbation theory expansion [5]. The DIP form factor $f(q_1^2, q_2^2)$, depends on the $q^2$ of each of the photons and has two free parameters, $\alpha_{DIP}$ and $\beta_{DIP}$:

$$f(q_1^2, q_2^2) = 1 + \alpha\left[\frac{q_1^2}{q_1^2 - m_\rho^2} + \frac{q_2^2}{q_2^2 - m_\rho^2}\right] + \beta\frac{q_1^2 q_2^2}{(q_1^2 - m_\rho^2)(q_2^2 - m_\rho^2)}$$

If one of the photons is real, $q_2^2$=0 and the term proportional to $\beta$ disappears. The DIP form factor is applicable to both $\gamma^*\gamma^*$ and $\gamma\gamma^*$ final states.

## KTEV DETECTOR AND DATA

KTeV took data in the rare-decay mode in 1997 and again in 1999-2000. The combined data set consisted of $6 \times 10^{11}$ $K_L$ decays, with a typical acceptance of 2-3%. Figure 2 shows a plan view of the KTeV detector, which combined powerful electromagnetic calorimetry with a precision spectrometer. Details of the detector can be found in [6].

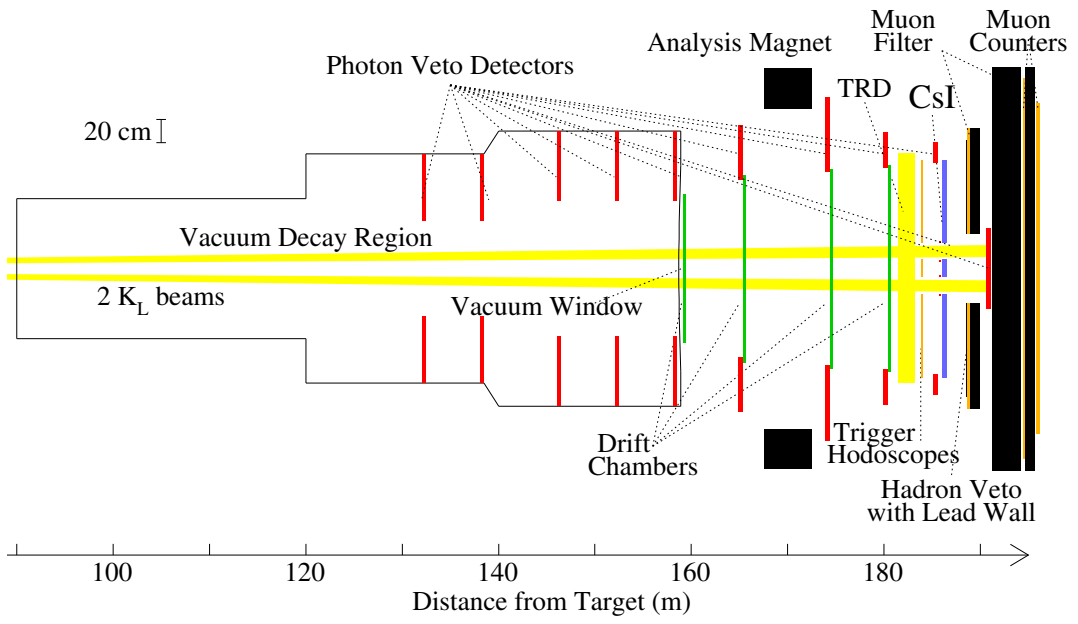

Figure 2. KTeV detector in E799 configuration

# ELECTROMAGNETIC MODES

## $K_L \to \mu^+\mu^-\gamma$

From the 97 data set we observe over 9000 candidate events for the decay $K_L \to \mu^+\mu^-\gamma$, with a background of 2.4% [7]. The mass peak is shown in Figure 3.

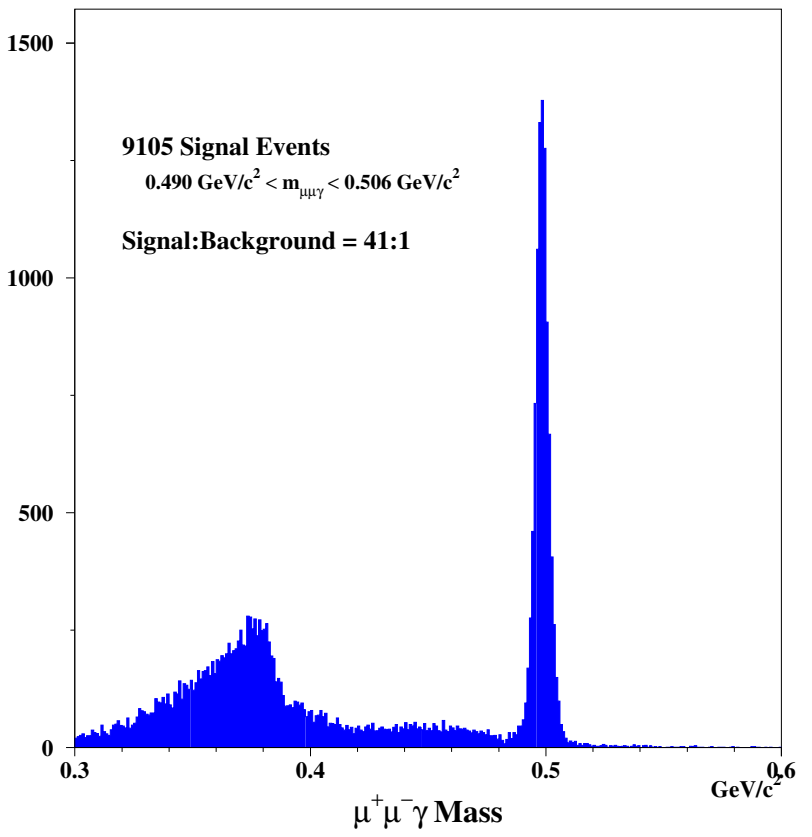

Figure 3. Mass distribution for $K_L \to \mu^+\mu^-\gamma$

From these data we measure the branching ratio and extract values for $\alpha_{K^*}$ and $\alpha_{DIP}$. The form factor fits are done by finding the minimum of the negative log-likelihood (NLL) function. As an example, figure 4 shows the NLL function for the determination of $\alpha_{K^*}$. Numerical values for the branching ratio and form factor parameters are given in Tables 1-3.

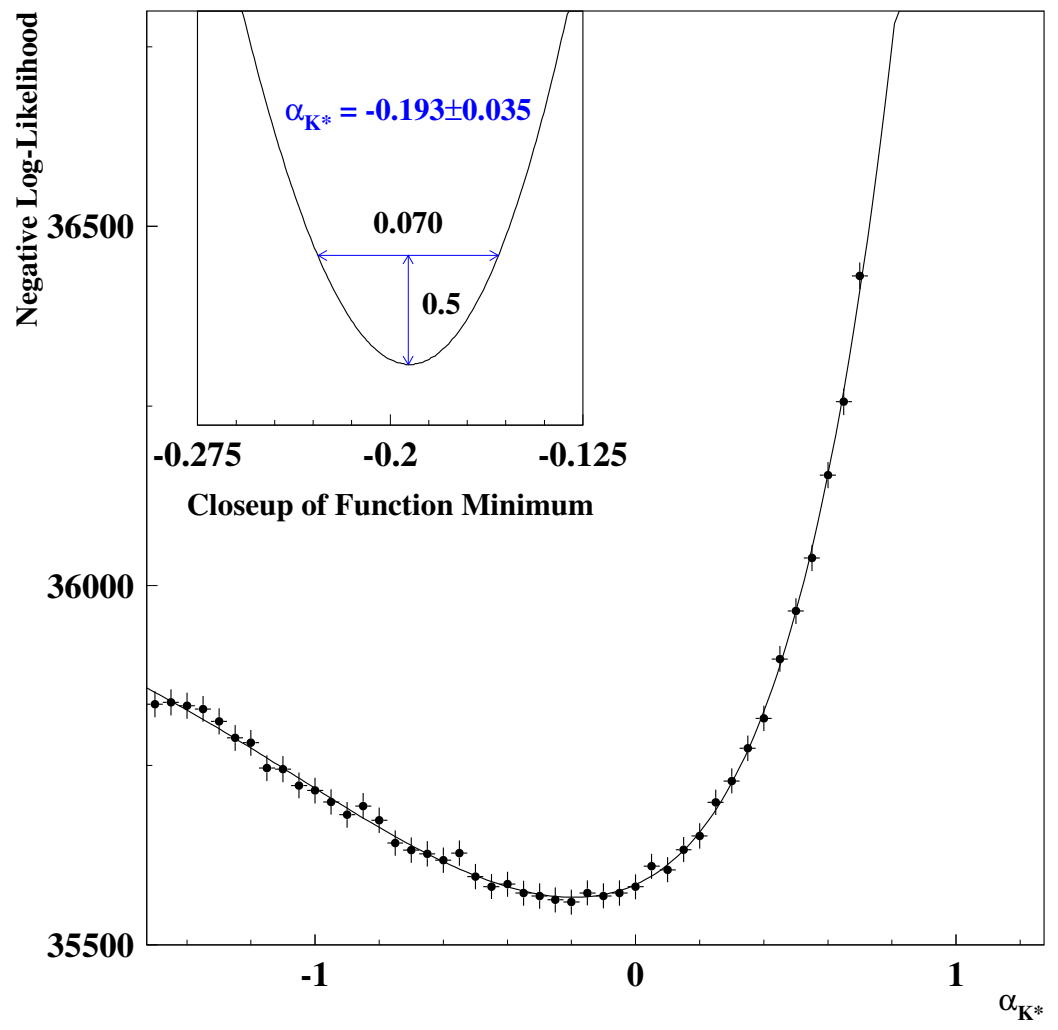

Figure 4. Negative log-likelihood function for the determination of $\alpha_{K^*}$ from $K_L \to \mu^+\mu^-\gamma$

## $K_L \to e^+e^-\gamma$

A closely related decay is $K_L \to e^+e^-\gamma$, but for this decay radiative corrections are important. The Colorado group has done a full set of radiative corrections to $\mathcal{O}(\alpha^3)$, for which details can be found in [8].

In the 97 dataset, we observe over 92,000 $K_L \to e^+e^-\gamma$ decays. Figure 5 shows the mass plot, and again the numbers for the branching ratio and form factor parameters are in the tables.

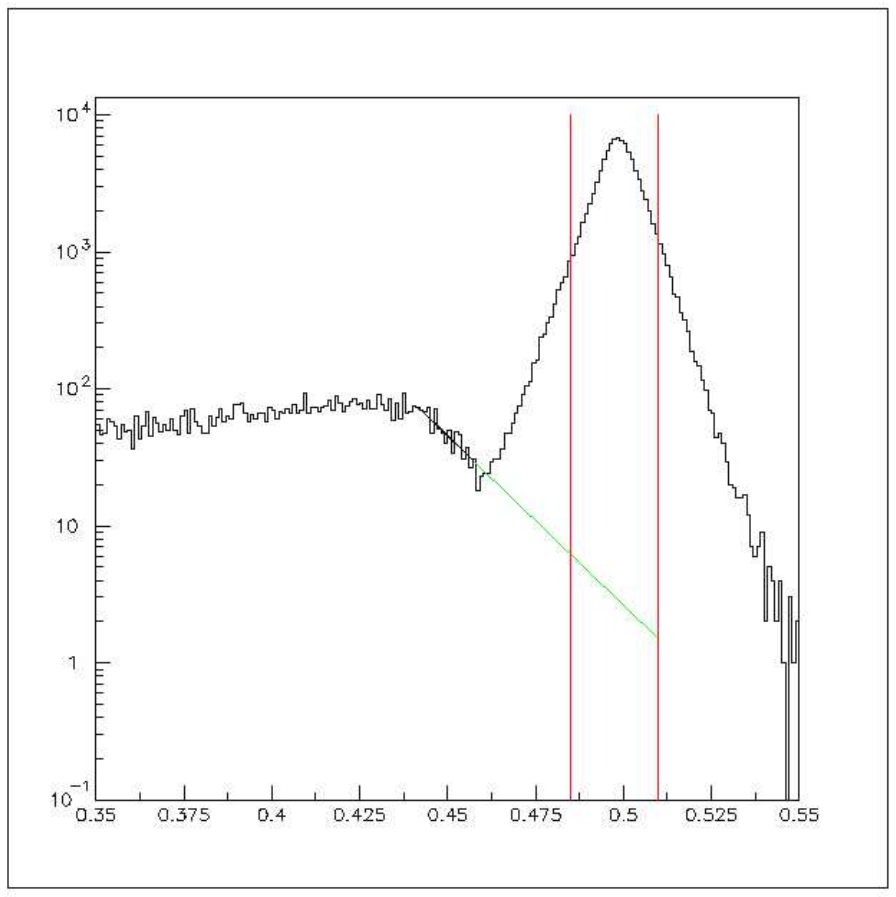

Figure 5. Mass distribution for $K_L \to e^+e^-\gamma$

Figure 6 compares three recent measurements of the BMS form factor parameter $\alpha_{K^*}$. The two KTeV measurements (from $ee\gamma$ and $\mu\mu\gamma$) agree with each other, but there is about a 3 standard deviation discrepancy between the KTeV and NA48 results which remains to be resolved.

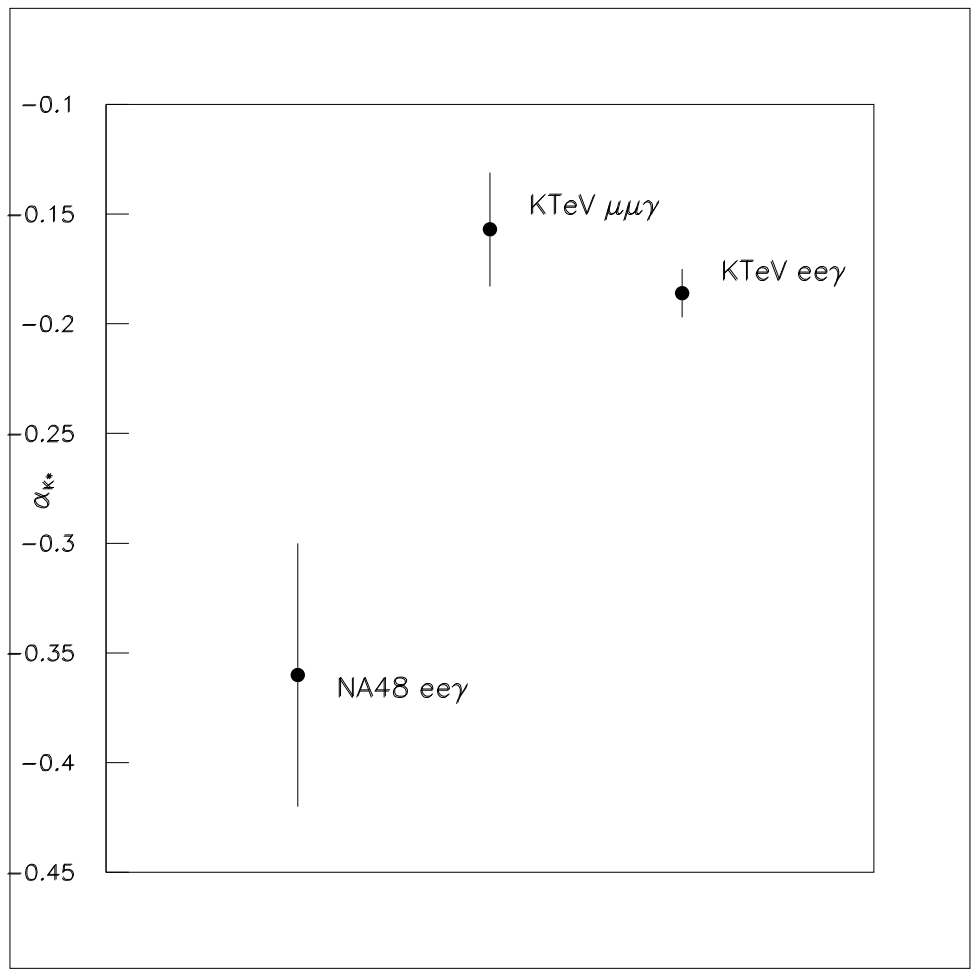

Figure 6. Recent measurements of $\alpha_{K^*}$

## $K_L \to e^+e^-\mu^+\mu^-$ *and* $K_L \to e^+e^-e^+e^-$

These modes are interesting because they have the potential to be sensitive to the form factor parameter $\beta_{DIP}$. Combining the 97 and 99 data, we find 132 candidate $K_L \to e^+e^-\mu^+\mu^-$ events, with an expected background of 0.8 events. [9] The improved branching ratio measurement is given in Table 1. Figure 7 shows the mass distribution for $K_L \to e^+e^-\mu^+\mu^-$ and the dominant background $K_L \to \pi^+\pi^-\pi^0_D$. We have also extracted a value for the DIP form factor parameter $\alpha_{DIP}$ from these data, but of course with much reduced statistical precision. At-tempts to extract $\beta_{DIP}$ did not yield meaningful results, as can be seen in figure 8. The log likelihood function has a very broad maximum, too broad to yield a determination of $\beta_{DIP}$.

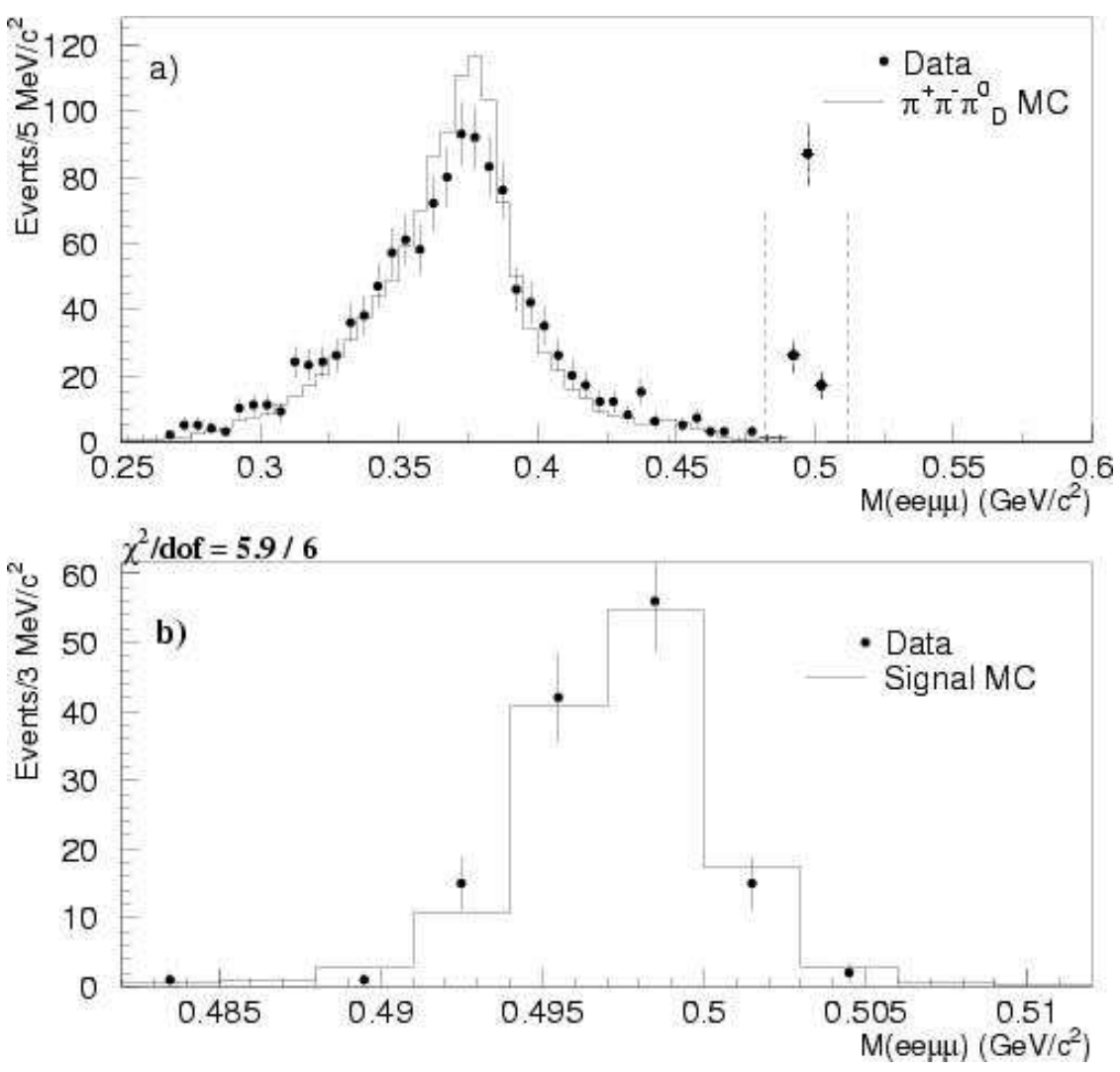

Figure 7. Mass distribution for $K_L \rightarrow e^+e^-\mu^+\mu^-$

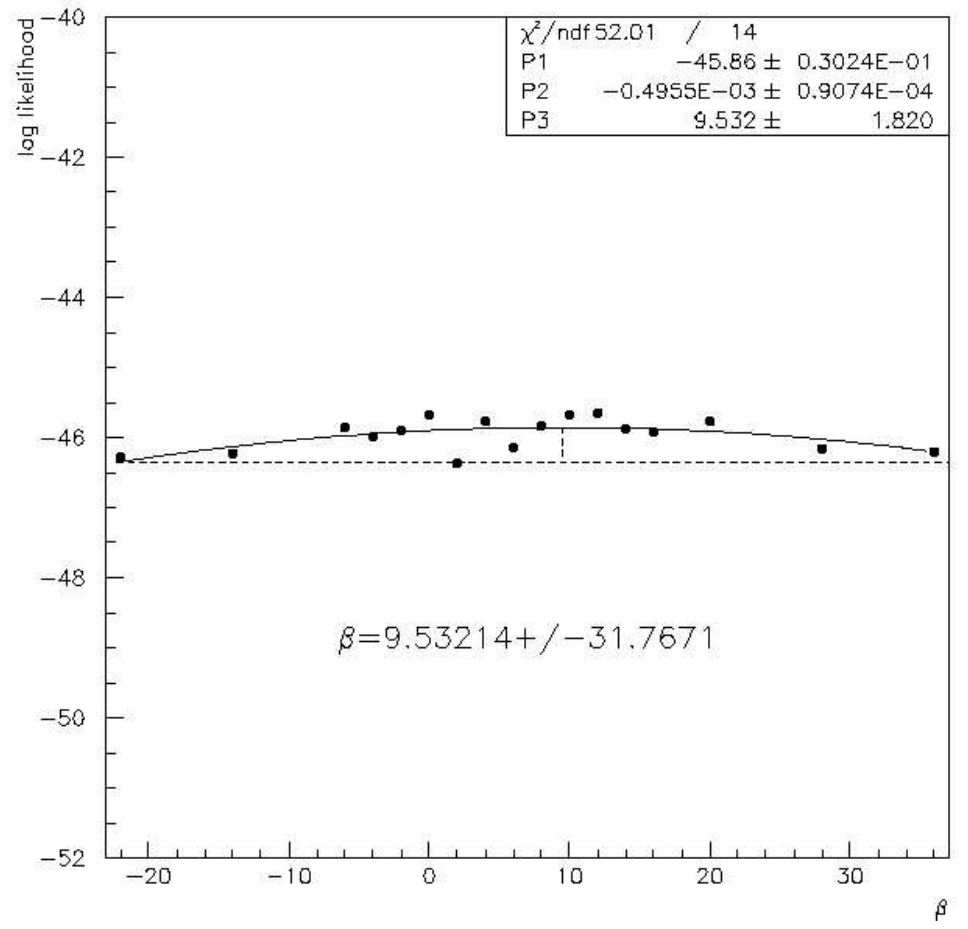

Figure 8. Log-likelihood function for $\beta_{DIP}$

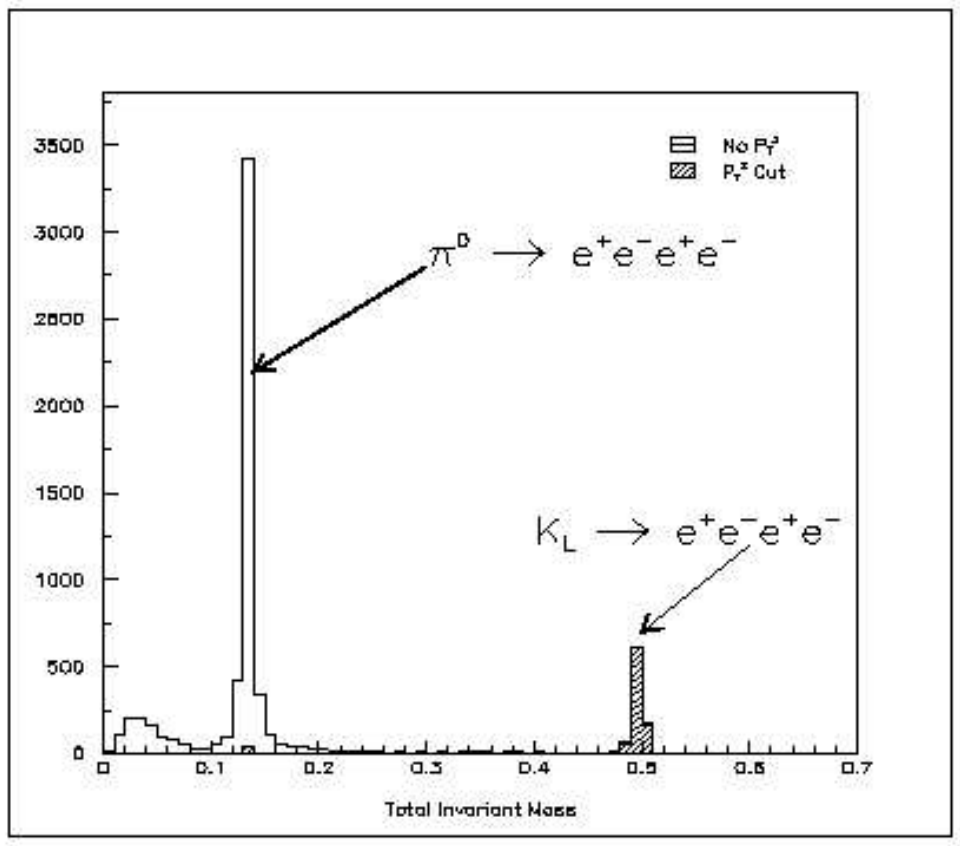

Figure 9. Mass distribution for $K_L \rightarrow e^+e^-e^+e^-$

For the $K_L \rightarrow e^+e^-e^+e^-$ decay we find 1056 events in the full KTeV data set, with an expected background of 6.5±0.3 events. The branching ratio result is given in Table 1, but again the statistics are too poor to allow determination of $\beta_{DIP}$. Figure 9 shows the mass distribution for $K_L \rightarrow e^+e^-e^+e^-$, along with the much larger mass peak for $\pi^0 \rightarrow e^+e^-e^+e^-$.

## $K_L \rightarrow \pi^0 e^+e^-$

This mode has long been of interest due to the expected direct-CP violating contribution. Interest in this mode has increased recently due to the observation by NA48 of the CP-conserving decay $K_S \rightarrow \pi^0 e^+e^-$, which determines the indirect-CP violating contribution to $K_L \rightarrow \pi^0 e^+e^-$ through $\epsilon$ [10]. Using the NA48 data, and additional theoretical arguments, Buchalla, D'Ambrosio and Isidori have recently [11] predicted a branching ratio for $K_L \rightarrow \pi^0 e^+e^-$ of $\sim 3 \times 10^{-11}$. From the 97 dataset, KTeV published a 90% CL limit of $5.1 \times 10^{-10}$[12]. Analysis of the 99 dataset has recently been completed, yielding an improved upper limit. Figure 10 plots the $ee\gamma\gamma$ mass ($M_{ee\gamma\gamma}$) vs. the $\gamma\gamma$ mass ($M_{\gamma\gamma}$) for the 99 dataset. In this plot, the vertex from the charged tracks is used to calculate the $M_{\gamma\gamma}$, while the neutral vertex (determined by assuming the $\pi^0$ mass for the two photons) is used to calculate $M_{ee\gamma\gamma}$. This procedure gives somewhat improved mass resolution for the signal mode. The major background, $K_L \rightarrow ee\gamma\gamma$, appears as a diagonal swath in this plot, from the upper left to the lower right. The small ellipse in the center is the signal region, while the larger box is the blind region, which is not examined until cuts are finalized. Figure 11 shows the data after all cuts except the cuts on the Greenlee kinematic variables $y_\gamma$ and $\theta_{min}$[13]. The variable $\theta_{min}$ is the minimum angle, as measured in the kaon rest frame, between any photon and any electron. The variable $y_{min}$ is the angle, as measured in the $\pi^0$ rest frame, between the kaon flight direction and the photon direction.

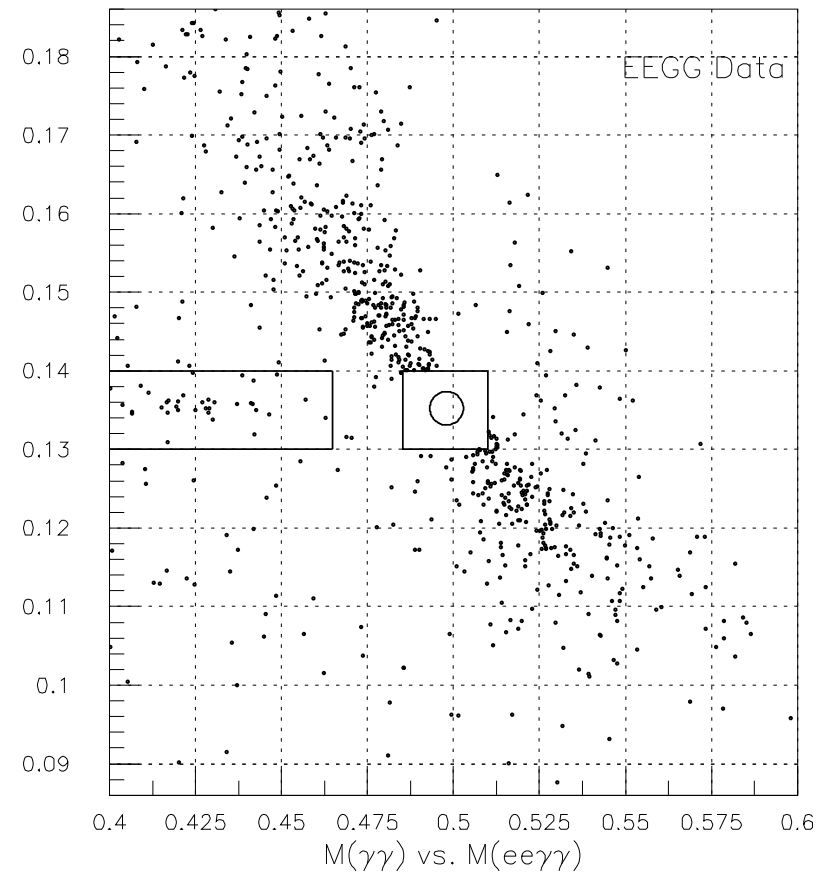

Figure 10. $M_{\gamma\gamma}$ vs. $M_{ee\gamma\gamma}$ for $K_L \rightarrow \pi^0 e^+e^-$, for the 99 dataset, before applying the cuts on the Greenlee variables.

These kinematic variables are very effective in removing much of the remaining background. Figure 11 shows the signal region and blind region after all cuts. The expected background in the signal ellipse is 0.99 ± 0.35 events, and we observe one event. We can set a 90% CL limit $BR(K_L \rightarrow \pi^0 e^+ e^-) < 3.5 \times 10^{-10}$ from the 99 data only, and a combined limit from the full KTeV dataset of $BR(K_L \rightarrow \pi^0 e^+ e^-) < 2.8 \times 10^{-10}$. Details of the analysis can be found in [14].

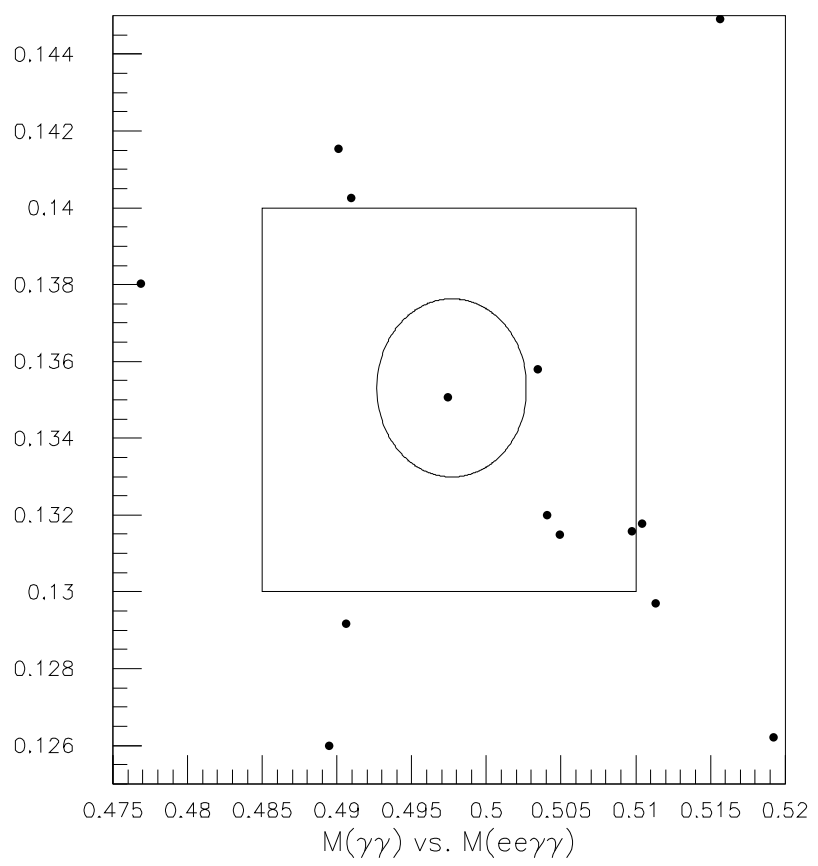

Figure 11. $M_{\gamma\gamma}$ vs. $M_{ee\gamma\gamma}$ for the 99 dataset, after all cuts have been applied. One event is present in the signal region, with 0.99 ± 0.35 background events expected.

## LEPTON FLAVOR VIOLATION SEARCHES

Many scenarios for physics beyond the Standard Model allow violation of lepton flavor number. Indeed, lepton flavor violation (LFV) has been observed in neutrino mixing, but it has yet to be observed in a decay process. KTeV has searched for lepton flavor violation in several channels. In these searches, we typically define a signal region in the two variables $p_t^2$ and kaon mass. $p_t^2$ is the vector sum of the transverse momenta of the final state particles with respect to the kaon flight direction. If all the decay products are observed and well-measured, $p_t^2$ will be close to zero. The signal box is chosen to contain about 90% of the events from the signal Monte Carlo. A study region about 100 times larger is defined in the $p_t^2-$Mass plane. The signal region is blind until all cuts have been selected, but the study region around the signal box is used to understand backgrounds.

One interesting LFV decay that KTeV might be sensitive to is $K_L \rightarrow \pi^0 \mu^\pm e^\mp$. We found three major backgrounds to this mode. $K_L \rightarrow \pi^+\pi^-\pi^0$ could fake the decay if one pion faked an electron and the other decayed or punched-through the steel to fake a muon. The excellent pion/electron separation eliminated most of this background, and the remaining events reconstructed at a mass well below the kaon mass, so this background did not encroach on the signal region. The second background is $K_L \rightarrow \pi^0\pi^\pm e^\mp \nu$, also known as $K_{e4}$. These decays were very similar to the signal mode, but due to the relatively small branching ratio and the fact that the $\pi^0 \mu e$ mass also reconstructs below the signal region, this turned out not a serious background.

The limiting background for the $K_L \rightarrow \pi^0 \mu^\pm e^\mp$ search was found to be $K_{e3}$ decays with pion decay or punch-though faking a muon, and with accidental photons faking a $\pi^0$. These events had distributions in both $p_t^2$ and kaon mass that were flat across the signal region, so we were unable to eliminate this background completely. Figure 12 shows the search region for the 99 data set, with the regions covered by the three backgrounds indicated. Although there are three events in the signal region, they are not distributed in a way consistent with signal, but rather are consistent with $K_{e3}$ plus accidental background. Even in the presence of this background, we can set a limit which is a factor of 20 better than previous measurements. The 90% CL limit for the full KTeV data set is 3.3 x $10^{-10}$.

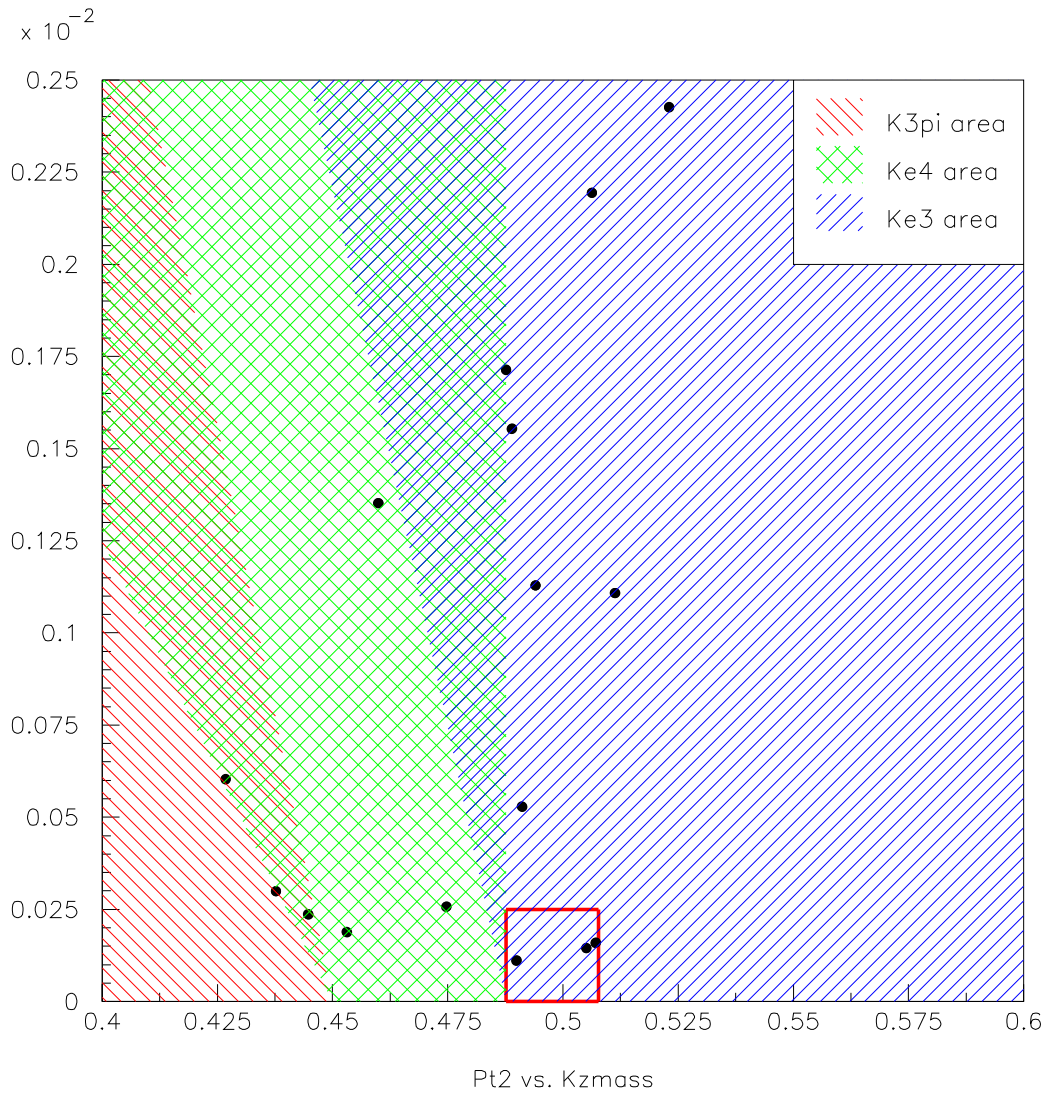

Figure 12. $p_t^2$ vs. kaon mass for the $K_L \rightarrow \pi^0 \mu^\pm e^\mp$ search from the 99 data set. All cuts have been applied.

Other LFV modes studied by KTeV include $\pi^0 \rightarrow \mu^\pm e^\mp$ with the $\pi^0$ tagged from $K_L \rightarrow \pi^0\pi^0\pi^0$. For this decay, the dominant background is $K_L \rightarrow \pi^0\pi^0\pi^0_D$ with one electron unobserved and an accidental muon. For the 97 data set, this background was small, and we observed no events in the signal region, and indeed no events in the much larger study region, as seen in figure 13. This allows us to set a 90% CL limit limit of 7.85 × $10^{-10}$. In addition, a search for the wrong sign combination in the $ee\mu\mu$ final state yielded no events, giving a 90% CL limit on the LFV decay $K_L \rightarrow e^\pm e^\pm \mu^\mp \mu^\mp$ of 4.12 × $10^{-11}$.

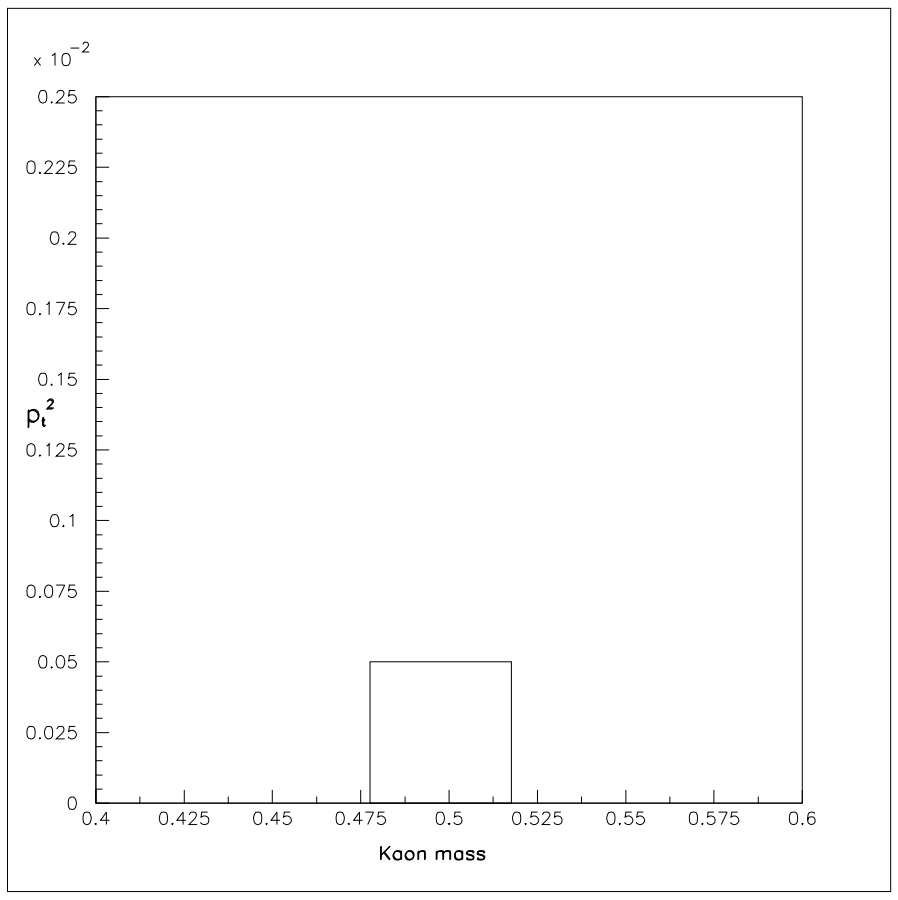

Figure 13. $p_t^2$ vs. kaon mass for the search $K_L \rightarrow \pi^0\pi^0\pi^0$ with $\pi^0 \rightarrow \mu^\pm e^\mp$ After all cuts, no events are present, either in the signal box or the larger search region.

## FUTURE PROSPECTS

Measurement of the form factor $\beta_{DIP}$ requires of order 100 to 1000 times more data than we were able to obtain in KTeV. Recording 100,000 $K_L \rightarrow e^+e^-\mu^+\mu^-$ decays is a formidable challenge, but one well worth the effort.

With a more reliable estimate for the $K_L \rightarrow \pi^0 e^+e^-$ branching ratio, it appears that observation of this decay may be within reach. Although the Greenlee background can never be eliminated, the expected branching ratio should allow observation of this decay over the $K_L \rightarrow e^+e^-\gamma\gamma$ background.

Lepton flavor violation in a decay process remains elusive. Neutrino mixing can lead to LFV decays, but only at a level far from the reach of existing or planned experiments. Many scenarios for physics beyond the Standard Model allow LFV decays, so these processes remain interesting as a possible window on new physics. In KTeV the LFV searches came along more or less for free. That is, a detector designed for rare decays and $Re(\epsilon'/\epsilon)$ measurements is well suited to the search for such forbidden processes. Such searches are a long shot, but with potentially large pay off.

It is clear that there remains a wealth of interesting physics to do in the kaon sector. In order to improve our level of knowledge, any future experiment must achieve unprecedented statistical precision. To set the scale, KTeV observed a total of $6 \times 10^{11}$ $K_L$ decays, with typical acceptances of 2-3%. A future experiment would need to have a (flux × acceptance) larger by at least a factor of 10.

Table 1: Recent KTeV results on branching ratios and limits. The first error quoted is statistical, and the second is systematic. The third error, when quoted, is an external systematic due to the uncertainty in the $\pi^0 \rightarrow \gamma e^+e^-$ branching ratio, which entered into the normalization.

| Decay mode | Branching Ratio |
|---|---|
| $K_L \rightarrow \mu^+\mu^-\gamma$ | $(3.62\pm0.04 \pm 0.08) \times 10^{-7}$ |
| $K_L \rightarrow e^+e^-\gamma$ | $(10.192 \pm 0.036 \pm 0.073 \pm 0.285) \times 10^{-6}$ |
| $K_L \rightarrow e^+e^-\mu^+\mu^-$ | $(2.69 \pm 0.24 \pm 0.12) \times 10^{-9}$ |
| $K_L \rightarrow e^+e^-e^+e^-$ | $(4.16 \pm 0.13 \pm 0.13 \pm 0.17) \times 10^{-8}$ |
| $K_L \rightarrow \pi^0 e^+e^-$ | $< 2.8 \times 10^{-10}$ (90% CL) |
| $K_L \rightarrow \pi^0\mu^\pm e^\mp$ | $< 3.3 \times 10^{-10}$ (90% CL) |
| $\pi^0 \rightarrow \mu^\pm e^\mp$ | $< 7.85 \times 10^{-10}$(90% CL) |
| $K_L \rightarrow e^\pm e^\pm \mu^\mp \mu^\mp$ | $< 4.12 \times 10^{-11}$ (90% CL) |

Table 2: KTeV results for the form factor parameter $\alpha_{K^*}$. First errors are statistical, second are systematic. If only one error is quoted it represents a combined error.

| Decay mode | $\alpha_{K^*}$ |
|---|---|
| $K_L \rightarrow \mu^+\mu^-\gamma$ | $-0.160^{+0.026}_{-0.028}$ |
| $K_L \rightarrow e^+e^-\gamma$ | $-0.186 \pm 0.011 \pm 0.009$ |

Table 3: KTeV results for the form factor parameter $\alpha_{DIP}$ First errors are statistical, second are systematic. If only one error is quoted it represents a combined error.

| Decay mode | $\alpha_{DIP}$ |
|---|---|
| $K_L \rightarrow \mu^+\mu^-\gamma$ | $-1.54 \pm 0.10$ |
| $K_L \rightarrow e^+e^-\gamma$ | $-1.63 \pm 0.038 \pm 0.028$ |
| $K_L \rightarrow e^+e^-\mu^+\mu^-$ | $-1.59 \pm 0.37$ |
| $K_L \rightarrow e^+e^-e^+e^-$ | $-1.08 \pm 0.41 \pm 0.13$ |

## REFERENCES

[1] L. Wolfenstein, Phys. Rev. Lett., **51**, 1945 (1983).

[2] G. Belanger and G. Q. Geng, Phys. Rev. **D43**, 140 (1991).

[3] L. Bergström, E. Massó, and P. Singer, Phys. Lett. **B249**, 141 (1990).

[4] . Buchalla and A. J. Buras, Nucl. Phys. **B412**, 106 (1994).

[5] G. D'Ambrosio, G. Isidori, and J. Portoles, Phys. Lett. **B423**, 385 (1998).

[6] A. Alavi-Harati et al., Phys. Rev. **D67**, 012005 (2003).

[7] A. Alavi-Harati et al, Phys. Rev. Lett. **87**, 071801 (2001).

[8] A. R. Barker, H. Huang, P. A. Toale, and J. Engle, Phys. Rev. **D67**, 033008 (2003).

[9] A. Alavi-Harati et al., Phys. Rev. Lett., **90** 141801 (2003).

[10] J. R. Batley et al, Phys Lett. **B576**, 43 (2003).

[11] G. Buchalla, G. D'Ambrosio, and G. Isidori, Nucl. Phys. **B672**, 387 (2003).

[12] A. Alavi-Harati et al., Phys. Rev. Lett., **86**, 397 (2001).

[13] H. Greenlee, Phys. Rev. **D42**, 3742 (1990).

[14] A. Alavi-Harati et al., **hep-ex/0309072**.